\documentclass[aip,apl,reprint,a4paper,floatfix,amsfonts,amsmath,amssymb,citeautoscript,groupedaddress]{revtex4-1}

\usepackage{mathptmx}
\usepackage[scaled=0.9]{helvet}
\usepackage[utf8]{inputenc}
\usepackage{graphicx}
\usepackage{upgreek}
\usepackage{textcomp}
\usepackage{sidecap}
\usepackage[
,textwidth=17.5cm
,textheight=23.5cm
,verbose
,dvips
]{geometry}
\usepackage{xspace}

\newcommand{\celsius}{$^{\circ}$C\xspace}
\newcommand{\dox}{$(D^{0},X_\text{A})$\xspace}
\newcommand{\sfx}{$(I_{1},X)$\xspace}
\newcommand{\idbx}{$(\text{IDB}^*,X)$\xspace}

\begin{document}

%Samples M2125 (on Ti foil)
	% M9960 (on Si)

\title{Molecular beam epitaxy of single crystalline GaN nanowires on a flexible Ti foil}

\author{Gabriele Calabrese}
\email{calabrese@pdi-berlin.de}
\author{Pierre Corfdir}
\author{Guanhui Gao}
\author{Carsten Pfüller}
\author{Achim Trampert}
\author{Oliver Brandt}
\author{Lutz Geelhaar}
\author{Sergio Fernández-Garrido}
\affiliation{Paul-Drude-Institut für Festkörperelektronik,
Hausvogteiplatz 5--7, 10117 Berlin, Germany}

\begin{abstract}
We demonstrate the self-assembled growth of vertically aligned GaN nanowire ensembles on a flexible Ti foil by plasma-assisted molecular beam epitaxy. The analysis of single nanowires by transmission electron microscopy reveals that they are single crystalline. Low-temperature photoluminescence spectroscopy demonstrates that, in comparison to standard GaN nanowires grown on Si, the nanowires prepared on the Ti foil exhibit a equivalent crystalline perfection, a higher density of basal-plane stacking faults, but a reduced density of inversion domain boundaries. The room-temperature photoluminescence spectrum of the nanowire ensemble is not influenced or degraded by the bending of the substrate. The present results pave the way for the fabrication of flexible optoelectronic devices based on GaN nanowires on metal foils.
\end{abstract}

\maketitle

The integration of electronic and optoelectronic devices on flexible substrates is motivated by the vision of novel and/or economically relevant applications.\citep{Kim2010,Dai2015} In this context, inorganic semiconductor nanowires (NWs) have recently emerged as promising candidates for flexible electronics and optoelectronics.\citep{Liu2015,Dai2015} Their large aspect ratio facilitates the expulsion of threading dislocation at the NW sidewalls, leading to their complete absence in the upper sections of well-developed NWs. This effect alleviates the requirement of lattice matching with the substrate and thus lifts the stringent constraints concerning the choice of the substrate material.\citep{Hersee2006,Hersee2011}

Indeed, the direct growth of semiconductor NW ensembles on flexible substrates has already been demonstrated for different material systems: ZnO NWs on polymer-based indium-tin-oxide coated substrates \citep{Nadarajah2008} and on paper, \citep{Manekkathodi2010} Ge NWs on plastic films \citep{Toko2015} as well as on Au coated Al foils \citep{Kumar2011} and Si NWs on stainless steel. \citep{Tsakalakos2007} However, the direct growth of inorganic semiconductors on flexible substrates is frequently hampered by either the low melting point of the organic materials used as substrate \citep{McAlpine2007} or the occurrence of interfacial reactions when semiconductors are directly grown on metallic surfaces.\citep{Bauer1990,Palmstrom1995} To overcome these issues and extend the possible range of material combinations, the wet etching of the substrate followed by NW embedding in a plastic film,\citep{Fan2009} and different lift-off techniques such as dry transfer \citep{Weisse2011,Madaria2010} and transfer printinting \citep{Lee2011,Triplett2014} have been developed. Using these methods, arrays of vertically oriented CdS \citep{Fan2009} and Si \citep{Weisse2012,McAlpine2007} NWs have been fabricated on flexible substrates. Regarding group-III nitrides, the material of choice for solid-state lighting and high-power electronics, the integration of GaN NWs on flexible substrates has been achieved either by transferring NW arrays grown on Si to polymer films \citep{Shi2015,Park2015} or by a lift-off process of NW ensembles grown on graphene coated Cu substrates.\citep{Park2014}

Metal foils are a particularly interesting type of substrate because they are not only flexible but also exhibit excellent electrical and thermal conductivities as well as a high optical reflectance. However, as pointed out above, the direct growth of inorganic semiconductors on metallic surfaces is challenging due to interfacial reactions. In the case of group-III nitrides, their growth as single crystalline NWs \citep{Woelz2015} and even the fabrication of NW-based light emitting diodes \citep{Sarwar2015,Zhao2016} have been demonstrated on Ti as well as Mo thick layers and bulk substrates. Therefore, the integration of GaN NW-based devices on metal foils seems to be within reach. Nevertheless, prior to device fabrication, the growth of single crystalline and well-oriented GaN NW ensembles on polycrystalline metal foils must be demonstrated.

%=====================================================================
%%Fig.1
\begin{figure*}
\includegraphics*[width=\textwidth]{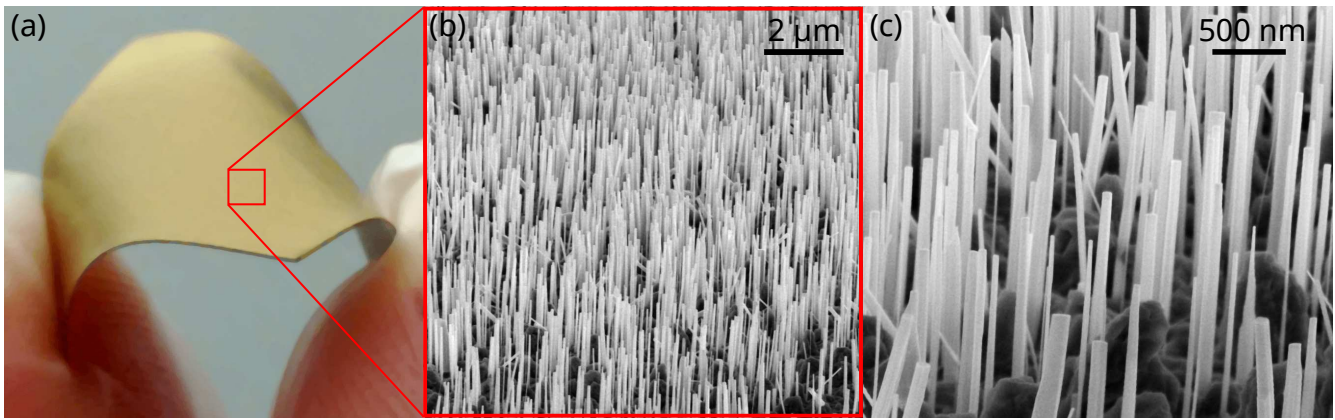} 
\caption{(a) Photograph of the Ti foil after NW growth demonstrating a high degree of flexibility. [(b), (c)] Scanning electron micrographs of the GaN NW ensemble grown on the Ti foil taken in bird’s eye view with (b) $8000\times$ and (c) $30000\times$ magnification. The red square in (a) is not to scale.}
\label{SEM}
\end{figure*}
%Samples:
%Figure 1: sample M2125
%=====================================================================

In this letter, we demonstrate the self-assembled growth of an ensemble of single crystalline, uncoalesced, and vertically aligned GaN NWs on a flexible Ti foil using plasma-assisted molecular beam epitaxy (PA-MBE). The structural and optical properties of the sample, investigated by transmission electron microscopy (TEM) and photoluminescence (PL) spectroscopy, are compared with those of a standard GaN NW ensemble prepared on a Si$(111)$ substrate. We find that both the structural perfection and the low temperature PL spectra of the NW ensemble prepared on the Ti foil and the Si substrate are fairly comparable. Furthermore, we do not observe any degradation of the luminescence upon bending the NW ensemble prepared on the foil down to a small curvature radius of 4~mm. Therefore, a GaN NW ensemble on a Ti foil is indeed a highly flexible system suitable for the realization of bendable GaN-based devices.   

The GaN NW ensemble presented here was grown by PA-MBE on a $127$~µm thick, $25\times25$~mm$^{2}$ large, polycrystalline and annealed Ti foil from Alpha Aesar with a purity of $99.99\%$. Using electron backscatter diffraction, we have estimated the grain size to be on the order of 50~nm, i.\,e., the foil is nanocrystalline. The PA-MBE system is equipped with a radio-frequency N$_{2}$ plasma source and a solid-source effusion cell for Ga. Prior to NW growth, the Ti foil was outgassed for $20$~minutes at 750\,\celsius in the preparation chamber of our MBE system and nitridated for $20$~minutes at 1000\,\celsius, as measured with a thermocouple placed in the substrate heater, to promote the formation of TiN on the surface. \citep{Woelz2015} The latter is a refractory material with metallic conductivity that forms an ohmic contact to GaN.\citep{Ruvimov1996,Luther1998} The growth of the GaN NW ensemble was performed at a substrate temperature of 730\,\celsius. The Ga and N fluxes, given as equivalent growth rate units of GaN films, were $5.3$ and $11.7$~nm/min, respectively. The total growth time was $4$~hours. Bright-field as well as high-resolution TEM and selected area electron diffraction (SAED) were carried out using a field emission JEOL-2100F microscope operated at $200$~kV. For these experiments, the NWs were dispersed on a carbon lacey grid. Continuous-wave photoluminescence (cw-PL) experiments were performed with the $325$~nm line of a He-Cd laser. The laser beam was focused to a $60$~µm spot, and the excitation density was $3$~mW\,cm$^{-2}$ ($3$~W\,cm$^{-2}$). For the experiments performed at $9$~K ($300$~K), the PL signal was dispersed by a spectrometer with $2400$~lines/mm ($600$~lines/mm) and detected by a charge-coupled device camera.

Figure~\ref{SEM}(a) presents a photograph of the Ti foil after GaN growth showing that the sample is indeed highly flexible. Bird's eye view scanning electron micrographs reveal the formation of a GaN NW ensemble [Figs.~\ref{SEM}(b) and \ref{SEM}(c)] with the majority of the NWs being vertically aligned. The large surface roughness observed in-between the NWs [see Fig.~\ref{SEM}(c)] is tentatively attributed to interfacial reactions between Ga and Ti at non-nitridated regions of the substrate surface. These reactions might also be at the origin of the misoriented NWs seen in Figs.~\ref{SEM}(b) and \ref{SEM}(c). Nevertheless, the density of misoriented NWs is low (between 1 and 2 orders of magnitude lower than that of vertical NWs) and they will be hardly contacted when fabricating devices based on axial NW heterostructures. Therefore, we believe they should not play a significant role on the final performance of future devices.

The statistical analysis of several cross-sectional and plane-view scanning electron micrographs (not shown here) with the help of the open-source software ImageJ, \citep{Abramoff2004} allows us to estimate the average values of the NW length and diameter as well as the NW number density. The corresponding values are 1.14~\textmu m, 70~nm, and $8.2\times10^{7}$~cm$^{-2}$, respectively. The NW density is, thus, more than one order of magnitude lower than the typical values observed in GaN NW ensembles prepared on Si$(111)$. \citep{Fernandez-Garrido2014} As result of their reduced number density, the NWs prepared on the Ti foil are almost completely uncoalesced (as deduced from plane view images not shown here). This result contrasts with the extremely high coalescence degrees (above $80\%$) observed in GaN NW ensembles prepared on other types of substrates such as Si or AlN.\citep{Brandt2014,Fernandez-Garrido2014} Since the coalescence of GaN NWs is a source of inhomogeneous strain \citep{Fernandez-Garrido2014} and results in the formation of non-radiative defects at the coalescence joints, \citep{Consonni2009} the possibility to fabricate NW ensembles free of coalescence on Ti foils represents a decisive advantage over other substrates.

\begin{figure}
\includegraphics*[width=\columnwidth]{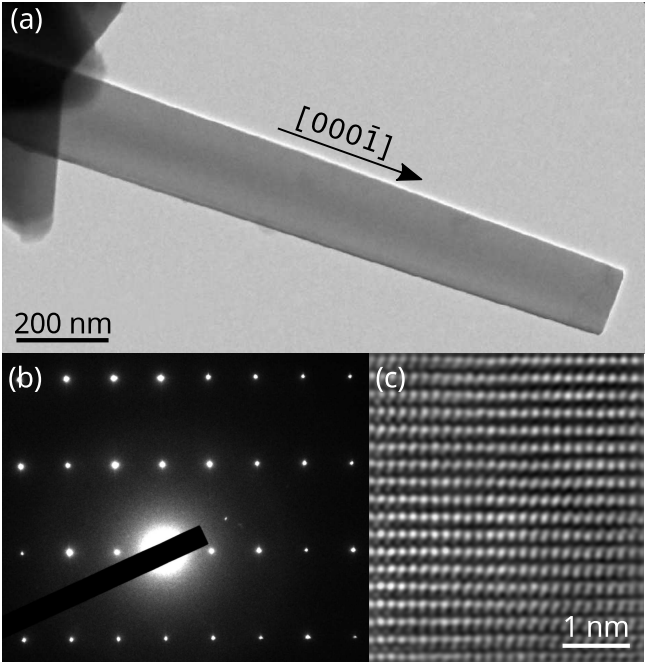} 
\caption{(a) Bright-field transmission electron micrograph of a single representative GaN NW grown on the Ti foil. The polarity of the NW, as indicated in the figure, has been determined from KOH etching experiments. (b) SAED pattern of the NW shown in (a). (c) Lattice image of a section of the NW shown in (a) taken along the $\langle 1\bar{1}00\rangle$ zone axis.}
\label{TEM}
\end{figure}

TEM of twenty individual GaN NWs was carried out to investigate their morphology and degree of crystallinity. Figure~\ref{TEM}(a) presents an exemplary bright-field transmission electron micrograph of a single NW. The image reveals a structurally uniform and dislocation-free NW without basal-plane stacking faults (we have only observed a single SF in one out of the twenty NWs we have investigated). The NW investigated in Figure ~\ref{TEM}(a) has a flat top facet, smooth sidewalls, and a diameter of about $150$~nm. Figure \ref{TEM}(b) displays the SAED pattern of the whole NW shown in Fig.~\ref{TEM}(a), confirming its single crystal nature and demonstrating that it crystallized in the wurtzite modification. The high-resolution micrograph shown in Fig.~\ref{TEM}(c) resolves the lattice of the wurtzite structure and attests to the high crystallinity of the GaN NW under investigation. These results suggest that the structural perfection of the GaN NWs grown on the Ti foil is comparable to that of uncoalesced NWs grown on Si$(111)$ substrates.\citep{Zettler2015}

The polarity of the GaN NWs was investigated on a macroscopic scale by KOH etching. \citep{Hestroffer2011,Romanyuk2015} For this purpose, several pieces of the sample were etched in a 5M KOH aqueous solution at $40^{\circ}$C for different times. The subsequent analysis of the samples by scanning electron microscopy (not shown here) revealed that, upon KOH exposure, all the investigated NWs become shorter. In addition, the smooth top facets seen in Fig.~\ref{TEM}(a) are no longer present after the chemical etching procedure. Instead, the NWs develop a pencil-like shape. NWs were not observed on the substrate after exposing the samples to KOH for 180 minutes. The evident etching of the GaN NWs in KOH demonstrates that they are N-polar, as commonly reported for self-assembled GaN NWs grown by PA-MBE.\citep{Fernandez-Garrido2012}

%=====================================================================
%%Fig.2
%Samples:
%Figure 1: sample M2125
%=====================================================================

%=====================================================================
%%Fig.3
\begin{figure}
\includegraphics*[width=\columnwidth]{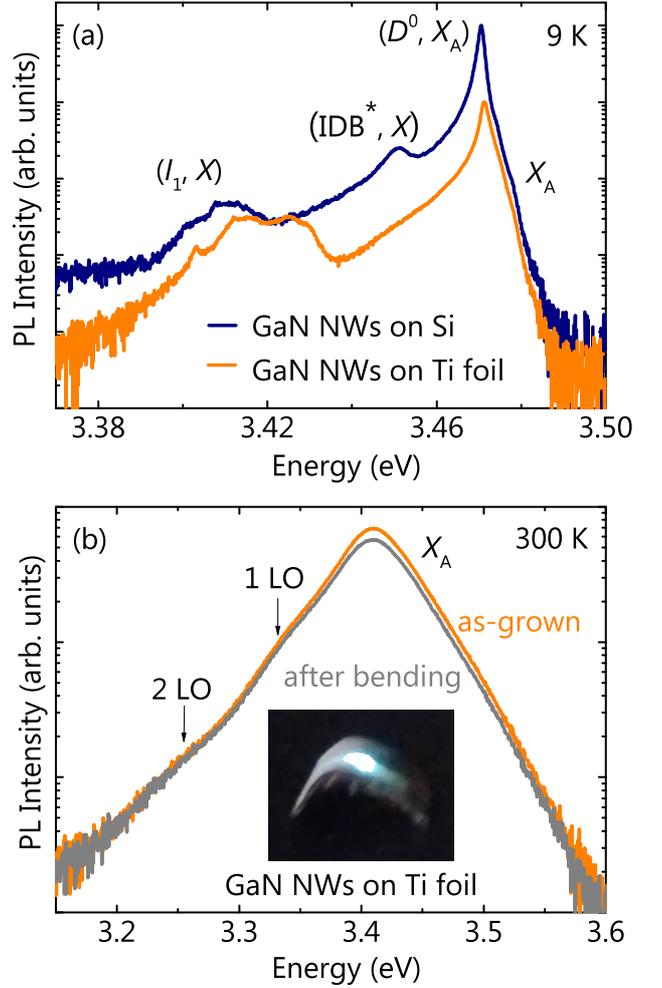} 
\caption{(a) Low temperature (9~K) PL spectrum of the GaN NW ensemble grown on the Ti foil. The corresponding spectrum of a GaN NW ensemble prepared on Si$(111)$ is included for comparison. The spectra have been normalized and shifted vertically for clarity. (b) Room-temperature PL spectra of the GaN NW ensemble grown on the Ti foil for an infinite and a convex curvature of the substrate with a 4~mm radius. The inset shows a photograph of the bent Ti foil taken during the acquisition of the experimental data.}
\label{PL}
\end{figure}
%Samples:
%Figure 1: sample M2125 and M9960
%=====================================================================

The structural perfection of the GaN NWs prepared on the Ti foil was additionally investigated by cw-PL spectroscopy. Figure~\ref{PL}(a) presents a near-band-edge low-temperature PL spectrum taken at 9~K on the nanowire ensemble under investigation. For comparison, we have also included the PL spectrum of a standard GaN NW ensemble prepared on Si$(111)$ under optimized growth conditions at 835\,\celsius.\cite{Zettler2015} The spectra are dominated by the recombination of excitons bound to neutral O donors [$(D^{0},X_\text{A})$] at $3.471$~eV. The transition energy corresponds to that of bulk GaN, revealing that GaN NWs are free of homogeneous strain.\citep{Calleja2000} The full-width-at-half-maximum (FWHM) of this transition, which is determined by the inhomogeneous strain \citep{Zettler2015} and the energy distribution of donors as result of their varying distances to the NW sidewalls \citep{Corfdir2014}, is $2.2$ and $1.5$~meV for the samples grown on Ti and Si, respectively. Therefore, the linewidth of the \dox transition for the NW ensemble on Ti is close to that of our optimized sample on Si and also comparable to other values reported in the literature.\citep{Calleja2000,SamGiao2013,Auzelle2015} 

For both samples, we additionally detect the recombination of free excitons ($X_\text{A}$) at $3.478$~eV, and the radiative recombination of excitons bound to $I_{1}$ basal-plane stacking faults SFs [$(I_{1},X_\text{A})$] between $3.40$ and $3.44$~eV. The intensity ratio between the \dox and the \sfx transitions is 50 and 300 for the NWs grown on Ti and Si, respectively. Consequently, the density of SFs seems to be slightly higher in the case of the GaN NWs grown on the Ti foil, most likely due to the significantly lower substrate temperature required to promote NW nucleation on TiN. In either case, since SFs in GaN NWs act as quantum wells confining excitons that recombine purely radiatively up to 60~K \citep{Corfdir2014b}, a very low density of SFs would be sufficient to explain the observed \dox/\sfx intensity ratio. This result is consistent with the fact that the majority of the NWs investigated by TEM are free of SFs. Finally, it is worth to note that the transition \idbx at 3.45 eV, which is clearly visible in the case of the NW ensemble prepared on Si, is much less pronounced for the NW ensemble grown on Ti. For GaN NWs, this transition has been recently identified to originate from inversion domain boundaries (IDBs).\citep{Auzelle2015} The absence of this transition suggests a reduced density of IDBs for growth on the Ti foil.

Room-temperature PL experiments were performed to analyze the effect of bending on the luminescence properties of the NW ensemble prepared on the Ti foil. Figure~\ref{PL}(b) presents the near-band-edge PL spectra prior and after bending the foil under a positive radius of curvature of $4$~mm. The figure also includes a photograph of the bent foil taken during the acquisition of the experimental data. As can be seen, the PL spectra prior to and after bending are essentially identical. The spectra are characterized by the free A exciton transition centered at $3.41$~eV with a FWHM of $64$~meV and its first and second-order longitudinal optical (LO) phonon replicas. These observations demonstrate that the GaN NWs are well anchored to the substrate and do not detach upon bending. 

In conclusion, we have demonstrated the self-assembled growth of single-crystalline, well-oriented, and uncoalesced N-polar wurtzite-GaN NWs on a flexible Ti foil. In comparison to GaN NW ensembles prepared on Si, the NWs grown on the foil exhibit a similar crystalline perfection, a lower degree of coalescence, a higher concentration of SFs, and a reduced density of IDBs. The NWs are found to be well anchored to the foil and do not degrade upon substrate bending. The present results pave the way for the realization of flexible GaN NW-based electronic and optoelectronic devices on metal foils.

We thank Carsten Stemmler, Michael Höricke and Hans-Peter Schönherr for their dedicated maintenance of the MBE system, Anne-Kathrin Bluhm for scanning electron microscopy, and Álvaro Guzmán for a critical reading of the manuscript. Financial support of this work by the Leibniz-Gemeinschaft under Grant SAW-2013-PDI-2 is gratefully acknowledged. P.\,C. also acknowledges funding from the Fonds National Suisse de la Recherche Scientifique through project 161032.

\bibliography{references}

\end{document}